\documentclass[hidelinks]{article}

\usepackage{arxiv}

\usepackage{orcid}

\usepackage[utf8]{inputenc} 
\usepackage[T1]{fontenc}    
\usepackage{hyperref}       
\usepackage{url}            
\usepackage{booktabs}       
\usepackage{amsfonts}       
\usepackage{nicefrac}       
\usepackage{microtype}      

\usepackage[english]{babel}
\usepackage{graphicx}
\usepackage[numbers,sort&compress]{natbib}

\newcommand\blfootnote[1]{%
  \begingroup
  \renewcommand\thefootnote{}\footnote{#1}%
  \addtocounter{footnote}{-1}%
  \endgroup
}

\title{Smart Monitoring: remote-monitoring technology of power, gas, and water consumption in Smart Cities}

\author{
  Sergey Surnov \orcidicon{0000-0002-3309-7326} \And Igor Bychkovskiy \orcidicon{0000-0001-6927-0031} \And Grigory Surnov \orcidicon{0000-0003-3408-3855} \And Sergey Krasnov \orcidicon{0000-0001-8325-7212}}

\begin{document}
\maketitle

\blfootnote{Sergey Surnov Ph.D., Igor Bychkovskiy, Grigory Surnov and Sergey Krasnov are with Divteh Ltd, Moscow, Russia.}

\blfootnote{\textit{Corresponding Author:} Sergey Surnov Ph.D., \href{mailto:s.surnov@divteh.ru}{\texttt{s.surnov@divteh.ru}}}

\begin{abstract}
This paper describes the remote-collection technology of detailed data (Smart Monitoring) on the consumption and quality of energy resources in public services. In this article, under “energy resources” (hereinafter referred to as resources) we outline electrical power, water (hot and cold), heat, and gas. Data on resource quality refer to the parameters characterizing the consumed resource. We also present an option of the data-acquisition system structure based on Smart Monitoring technology. Particular attention is paid to security in the system and the centralized management of its elements. The data flow in such system carries information about the behavior of energy consumers and the household equipment they use. Data on energy consumption for billing purposes in such a system is just one of many kinds, and not the most important feature. The development of Smart Monitoring technology is aimed at developing the market of IT services and mass services based on analysis of collected detailed data on energy-resource consumption.
\end{abstract}


\section{Introduction}
The modern trend in the creation of data-collection systems on resource consumption (hereinafter referred to as Systems) is the remote collection of detailed consumption data. The main source of data in a System are resource meters. Meters that remotely transmit detailed data on resource consumption are now called “smart.” Systems with smart meters belong to the field of the Internet of Things (IoT) and have their own characteristics. For example, meters do not need to transfer data to each other, and the data transferred by the meters do not need to immediately be responded to. The population's consumption of resources is uneven---by analogy with wired telephony, hours of the highest load can be distinguished.

Remote data collection conditions following:

\begin{enumerate}
\item The meter becomes part of the System with a monitoring center where data on resource consumption are remotely collected.
\item Meter design should provide for embedding them in such a system.
\item There can be many, or even more, meters in the System; they can be distributed over a large area. This requires special attention to System maintenance and security \cite{l18,s17}.
\end{enumerate}

When designing a System, it is necessary to answer two closely related questions: what kind of data network from the meter to the Monitoring Center should be used, and which data do we want to collect?

The first question is currently receiving a great deal of attention \cite{l18}: international alliances have been created, and several competing solutions are being developed, the leading among them are Long Range Wide-Area Networks (LoRaWAN) and Narrow Band IoT (NB IoT).

The answer to the second question is not as obvious as it seems at first glance. Our paper is devoted to discussing this very question. We emphasize that the article is dedicated to solving the problem of the remote collection of as data that are as detailed as possible on resource consumption by end users. The problem of interpreting these data by extracting the information contained inside them for further use lies outside the scope of our article. We hope that the data collected by our technology will become a good basis for specialists in energy-saving, machine-learning, and artificial intelligence in developing various kinds of applications.

\section{Which Data Do We Want to Collect?}

In general, data that are known to the resource meter and which it can transmit can be divided into two groups:

\begin{enumerate}
\item Data on the state of the meter itself and on unauthorized interference with its operation.
\item Data on the quantity and quality of consumed resources.
\end{enumerate}

The first group of data is important for System operation. These data carry information about the state of the equipment of which the System consists, and on attempts to interfere with the System. They can also be used to upgrade the System.

The issue with the second group of data is more complicated. In order to simplify it, we can draw a few fairly reasonable assumptions.

We get the minimum amount of information when collecting data only at the end of the month, and the maximum when collecting data in real time. It is not a mistake to assume that the amount of information contained in the collected data is some function of time interval $\Delta t$, through which we poll meters:

$I=f(\Delta t),$

$ I_{max}=f(\Delta t\rightarrow 0),\ I_{min}=f(\Delta t=1$ month$)$.

The dependence of the amount of information on $\Delta t$ is shown in Figure~1.

\begin{figure}[ht]
\centering
\includegraphics[width=.6\linewidth]{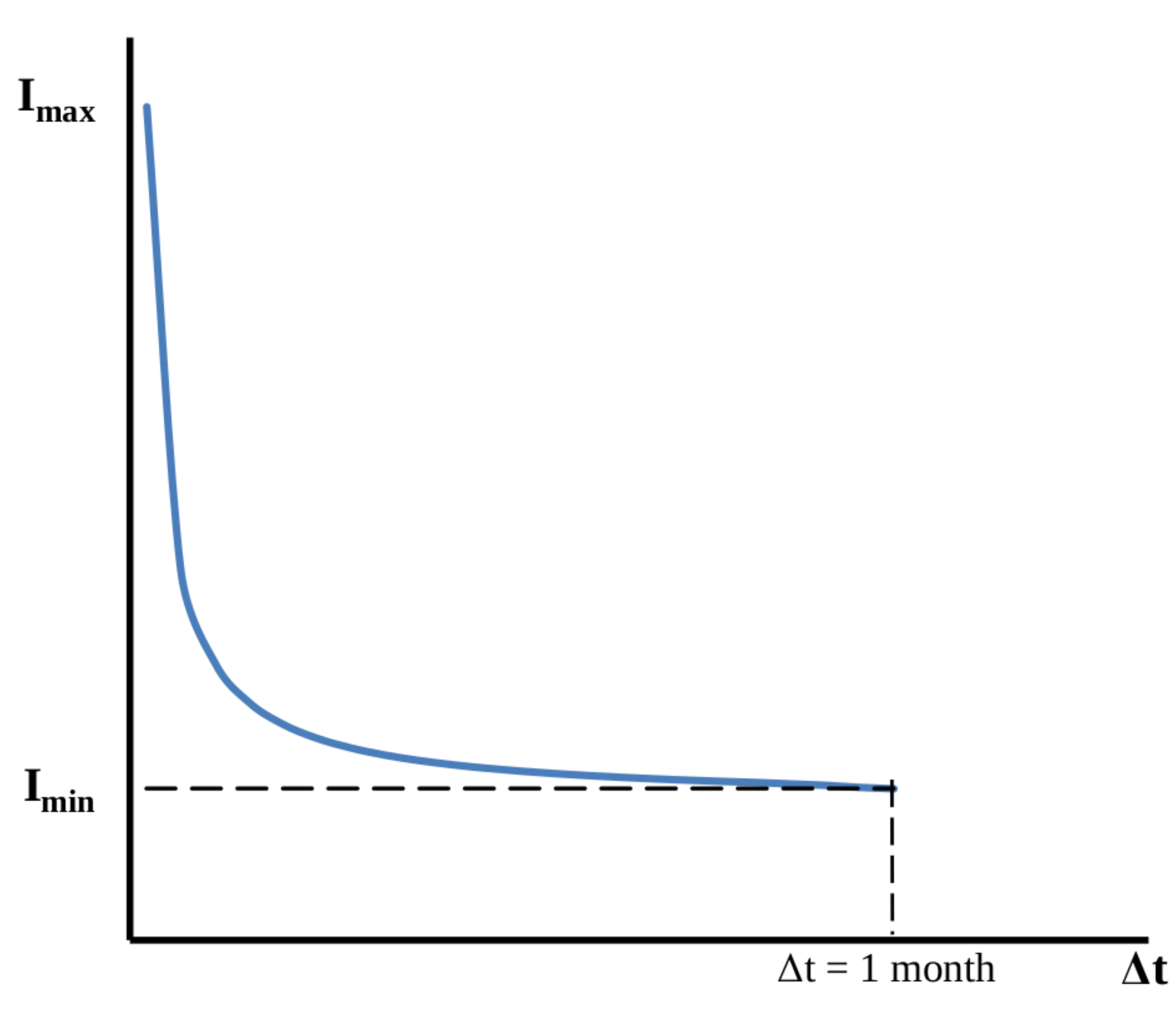}
\caption{Dependence of amount of information on $\Delta t$.}
\end{figure}

If $\Delta t = 1$ month, then the obtained data can most likely be used only for billing, but we cannot find out how and what quality of resources were consumed during the month.

Almost all major meter manufacturers also produce Remote Systems to obtain their readings. The goal is to provide a convenient tool for settlements between suppliers and resource consumers, as well as to obtain some kind of operational data. Readings are retrieved periodically at a predetermined (long enough) time interval $\Delta t$ (month, week, day) \cite{sr14}. The collected data in this case carry some information, but it is clearly insufficient to draw conclusions about consumer behavior or the equipment used by them. Such Systems mainly solve the problems of the resource provider, but not the consumer.

The emergence of nonintrusive appliance load monitoring (NILM) technology \cite{h93} and many works based on its application have shown how much information that detailed data carry on electricity consumption. It turned out that they contain information about the behavior of consumers, applied household equipment, and the power consumption by each device \cite{wz12,al17,aasssr17,h18}. There are new applications, sometimes unexpected, that rely on analysis of detailed data, for example, monitoring the behavior of single elderly people and diagnosing diseases \cite{au17}. Increasingly complex methods of analyzing detailed data are being used, for example, machine learning and neural networks \cite{kk15,p15,aasssr17}. However, the massive use of NILM meters in public services is constrained by their complexity and cost \cite{bkb10,vde10}. In this regard, attempts are being made to lower the polling frequency in NILM meters \cite{m14}.

There is every reason to believe that detailed data on the consumption of other resource types by populations (water, heat, and gas) also carry a lot of information. A joint analysis of detailed data on the consumption of all types of resources by a single consumer (apartment or private house) can reveal much more information than analysis of one of the resources.

For electricity, the level of data detail is very high, close to real time. In practice, it is very difficult to achieve the same level of data detail on the consumption of water, heat, and gas. Attainable polling interval $\Delta t$ for these meters does not provide an acceptable level of detail at which, for example, the devices used by the consumer can be identified. The polling interval cannot be significantly reduced due to the discharge of autonomous batteries in meters and the overload of data-transmission channels. The system may become sensitive to the number of meters therein. The problem of the excessive loading of data-transmission channels is, however, the case of when using NILM meters.

Modern remote data-collection systems operate on the principle of meter readings after a certain predetermined time interval. An event in such systems is the passage of a predetermined time interval $\Delta t$, a reaction to an event, that is, meter readings. We call such systems time-interval systems (Ti systems).

Consumption of resources by the population during the day is very uneven. However, in Ti systems, data are collected regardless of whether resources are being consumed or not. Accordingly, the channels of communication are also loaded. This is a significant disadvantage of Ti systems.

\section{Smart Monitoring Technology}
The desire to avoid energy costs and loading communication channels when no resources are being consumed, to achieve an acceptable level of data detail, improve security in the data-collection system, and simplify the meters led to the creation of Smart Monitoring technology. The technology is based on the following principles:

\begin{enumerate}
\item No consumption, no data: data from the meters are transmitted only when resources are consumed.
\item Registering resource-consumption time.
\item No access to the meter from the outside: unauthorized access to a huge number of meters in city conditions and their malfunction may have the most unpleasant consequences.
\end{enumerate}

In Smart Monitoring technologies, for each meter, the $\Delta R$ value of the amount of the resource passed through the meter is specified. An event is the passage of a specified amount of resource $\Delta R$ through the meter, and the response to the event is the transfer of data by the meter.

We call such systems resource-interval systems (Ri systems). Initial value $\Delta R$ for each resource is given on the basis of some “reasonable" considerations. On the one hand, it is desirable to make $\Delta R$ as low as possible, but in practice it would be “unwise”, for example, to track the consumption of every milliliter of water in the city.

In the future, $\Delta R$ can be adjusted according to operation results, taking into account actual limitations. For example, taking into account the actual (not calculated) discharge of autonomous batteries, which in Ri systems are a function of $\Delta R$ and the amount of actually consumed resource.

Resource consumption in an Ri system is depicted in Figure~2.

\begin{figure}[ht]
\centering
\includegraphics[width=.8\linewidth]{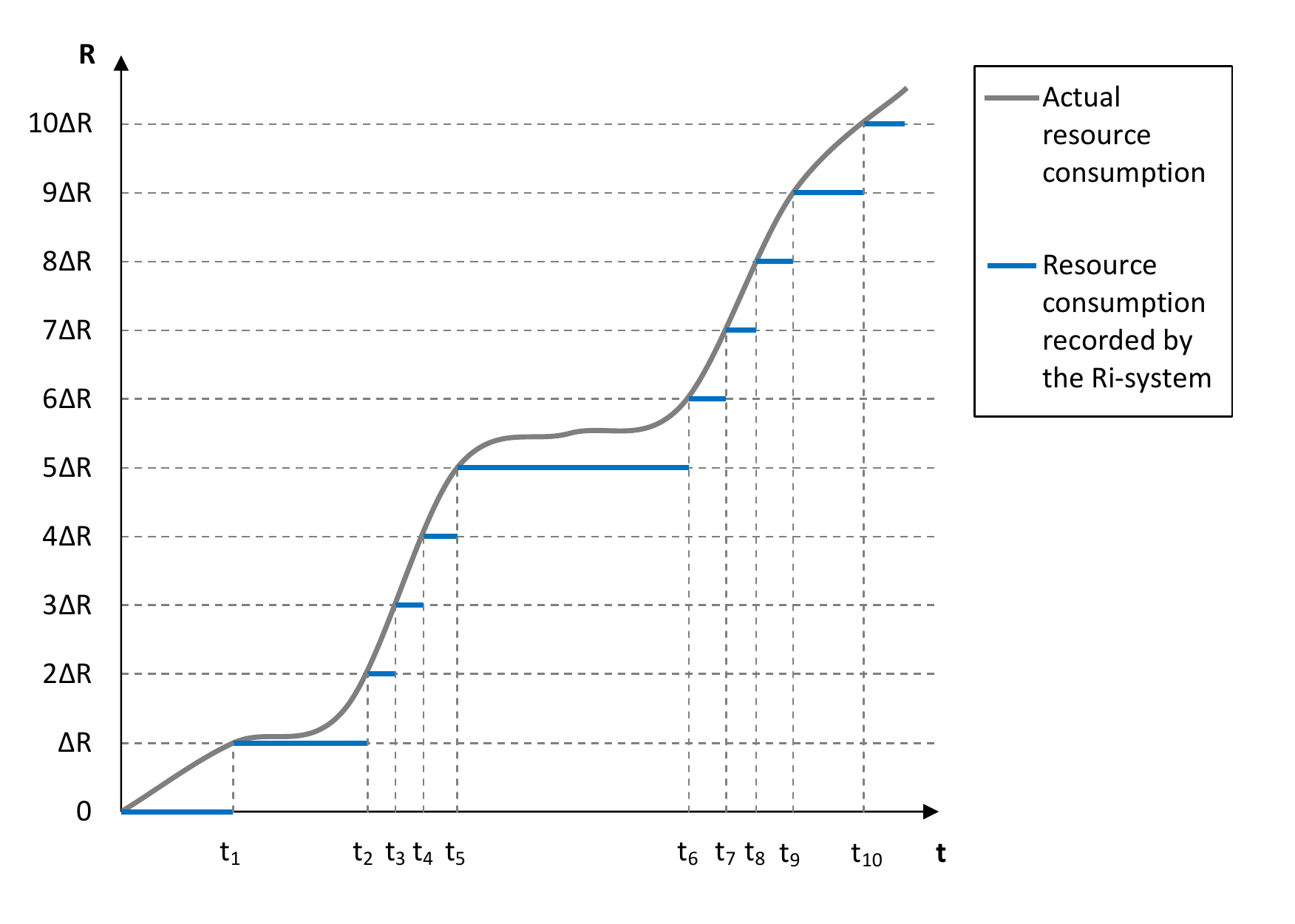}
\caption{Resource consumption recorded by a resource-interval system (Ri system).}
\end{figure}

One of many possible structures of an Ri system that implements Smart Monitoring technology is shown in Figure~3.

\begin{figure}[ht]
\centering
\includegraphics[width=.8\linewidth]{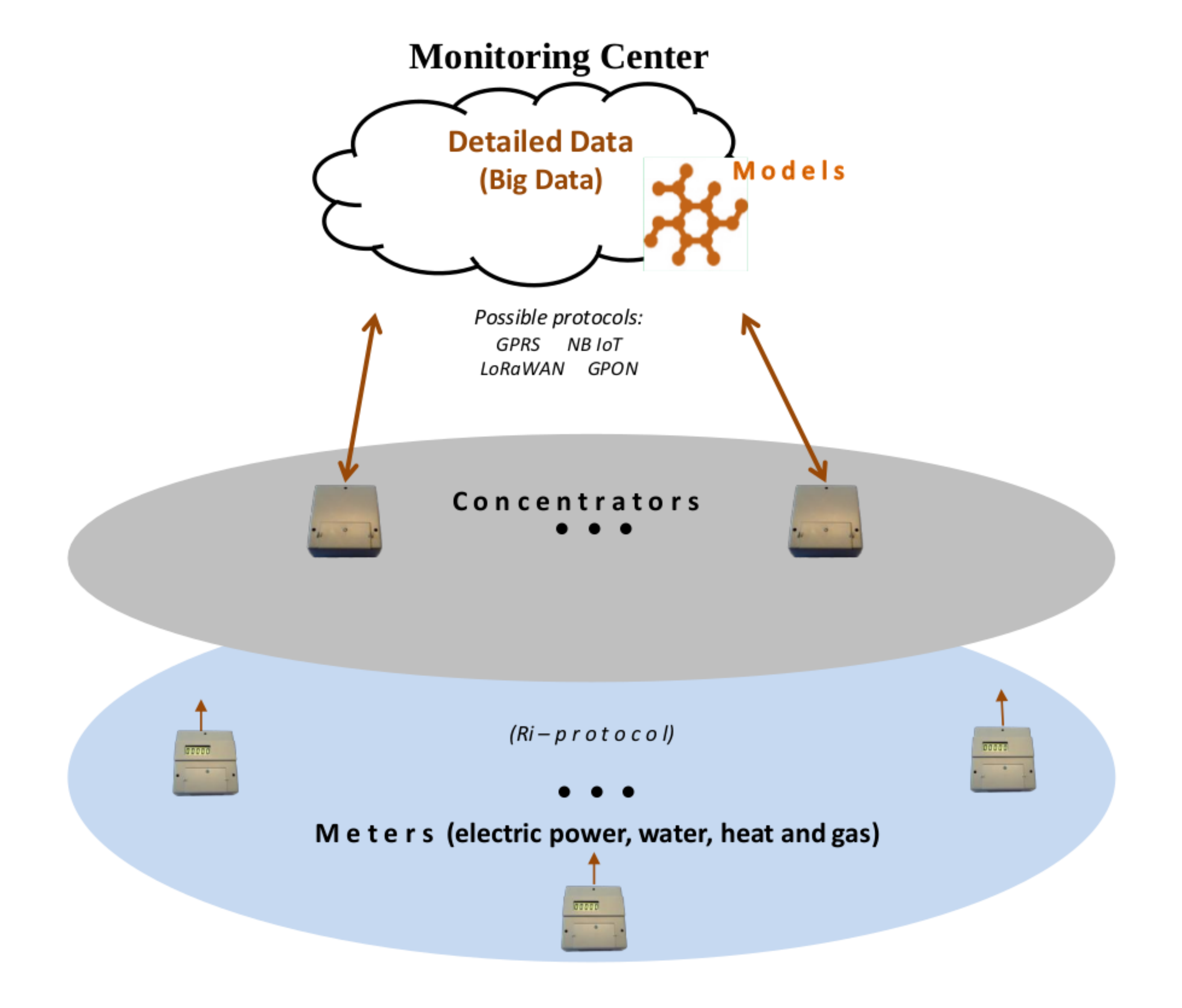}
\caption{Possible structure of Ri system that implements Smart Monitoring technology.}
\end{figure}

There are three levels in an Ri system: the meter, the concentrator, and the Monitoring Center levels.

The essence of the Ri protocol (communication between meters and concentrators) is as follows.

As soon as a specified amount of resource $\Delta R$ has passed through the meter in an Ri system, the meter transmits its number, session number, resource quality, and its state data to all the surrounding concentrators. Communication is one-way, from the meter to the concentrator. If the resource is not consumed for a long time, the meter transmits data about its state after a certain time interval (for example, after 24 hours).

Concentrators within the radio-visibility range of the transmitting meter receive the data transmitted to them, and add the timestamp of when it was received to the received message, as well as information about their state. Then, all data are transmitted to the Monitoring Center. Time synchronization in the concentrator and in its operation algorithms is performed from the Monitoring Center. Concentrators can be located outside apartments in an apartment building and are provided with main power. Concentrators 'do not distinguish' types of meters by type of resources. The same message transmitted by a meter can be received by several concentrators.

Data processing is done in the Monitoring Center. It also contains all the information about the meters and concentrators installed in the Ri system.

Such architecture of an Ri system requires a unique numbering of meters and concentrators. To do this, the identification of things (IDoT) in Digital Object Architecture (DOA) technology can be used \cite{kw06,sfi16}.

Ri systems are essentially asynchronous systems where meters are always communication initiators.

At first glance, we can get away without the concentrator level. However, the introduction of this level solves two key problems at once:

\begin{enumerate}
\item Utmost simplification of meters as the most massive device in the Ri system. The complexity of equipment and software in an Ri system increases from the meter to the Monitoring Center.
\item Prevention of unauthorized access to the meters.
\end{enumerate}

The problem of concentrator vulnerability remains, but it is not as critical as in the case of meters. In addition, in Smart Monitoring technology, the ratio of the concentrator number to the meter number in a standard apartment building is no more than 1:50 (when installing meters for all types of resources), and concentrators that are in the radio-visibility range of the same meter duplicate each other.

The interface between concentrators and Monitoring Center can be any kind, for example, General Packet Radio Service (GPRS), Gigabit-capable Passive Optical Network (GPON), NB IoT, or LoRaWAN. The choice of this interface does not affect the interaction of meters with hubs.

The proposed solution leads to data-collection technology that implements the following formula:

\textit{Smart Monitoring = remote data collection + acceptable data detailing + Cloud service + per apartment accounting + remote system monitoring and control + acceptable equipment cost + acceptable security level.}

As is rightly noted in \cite{l18}, besides security, the most important requirement for an IoT system is to maintain performance while increasing its scale. The Ri system is easily scaled to the territory of any size. Calculations show that it is impossible to “overload” communication channels with an increase in the number of meters, since the number of meters per unit area of the territory are limited by the capabilities of residential development. Even simultaneous consumption of the maximum possible amount of all types of resources by all customers could not disable an Ri system.

In an Ri system, there may be accidental losses of individual messages (for example, due to interference) even despite duplication of message reception by concentrators. However, the numbers of lost messages are known in the Monitoring Center, and the amount of resource consumed is easily restored. In some cases, by using the profile of the consumer (consumption traditions), the transmission time of lost messages can be restored. Even if this fails, however, the loss of individual messages is not critical for our purposes.

The continuous flow of detailed data (Big Data) from consumption meters, resource quality, equipment condition, and various abnormal situations contains information about the behavior of the resource consumer, the equipment used by it, and the state of the Ri system itself. This information can be used for billing, energy saving, optimization of the structure of the Ri-system itself, etc.

An application for the invention of the Smart Monitoring technology has been filed in the United States \cite{uspat18}.

\section{Experimental Data-Collection System}
Using Smart Monitoring technology, an experimental Ri system for collecting data on the consumption of cold and hot water, electricity, and heat energy, was deployed. A small batch of apartment meters with a transmitter power of 8 mW was released. The cost of a tachometer water meter turned out to be approximately equal to the cost of a traditional apartment tachometer water meter with a pulse output. Value $\Delta R$ for the meters was set as follows: cold and hot water, 100 ml; electric power, 10 Wh; heat energy, 5 kcal. Two interfaces were implemented between the hubs and the Cloud: GPRS and GPON. Operation of the Ri system confirmed the efficiency of the adopted solutions.

\section{Consequences of Using Smart Monitoring technology}
Noted are some interesting consequences of using Smart Monitoring technology (the list is not exhaustive):

\begin{enumerate}
\item Time for complete discharge of autonomous batteries in different meters varies because a different resource amount passes through different meters, and the battery discharge is its function.
\item During meter life, one can observe the “deviation” of its parameters, for example, an increase in error. To minimize such effects, it is possible to either calibrate the meter or adjust its readings in the Monitoring Center and do so without calibration.
\item Any other equipment operating on the principles of Smart Monitoring technology, for example, temperature sensors in each room, water-leakage sensors, gas analyzers, and security sensors can easily be added to the Ri system.
\end{enumerate}

\section{Conclusion}
The suggested Smart Monitoring technology is focused on collecting data on resource consumption in public services and heavily relies on its features. It does not aim to collect data in real time, as is possible with NILM technology. However, it makes it possible to significantly simplify meter design, achieve quite an acceptable level of data detailing, unload communication channels, and increase the level of security. The mere possibility of joint analysis of the consumption of all resource types in an apartment that this technology offers compensates, to a large extent, for its shortcomings.

We hope that the technology we developed allows us to take a step forward in obtaining data on the behavior of resource consumers in the public services. This, in turn, would contribute to the development of the market for IT services, and mass services for end users in the control and efficient use of electricity, heat, water, and gas.

\section*{Acknowledgments}
The authors are grateful to Intel Corp. (USA) that provided its Cloud for the experiments, the final equipment of the Ri system subscriber, and their very useful advice, as well as Andrey Mikhailovich Turlikov, Doctor of Engineering Sciences, Head of the Information and Communication Systems Department (State University of Aerospace Instrumentation, Saint Petersburg) who suggested a number of valuable comments on the data-transmission system.


\begin{thebibliography}{99}
\bibitem{l18}
Perry Lea. “Internet of Things for Architects”. Publisher: Packt Publishing. January 2018. ISBN:~9781788470599.
\bibitem{s17}
Siemens. “Smart meters - safety comes first!”. Sep. 08, 2017. Accessed: Mar. 29, 2019. [Online]. Available: \url{http://w5.siemens.com/web/ua/ru/news_press/news/2017/Pages/smart_metering_infrastruktur.aspx}
\bibitem{sr14}
B. Subhash, V. Rajagopal, “Overview of smart metering system in Smart Grid scenario”, IEEE Conferences: Power and Energy Systems: Towards Sustainable Energy, pp. 1-6, 2014, doi: 10.1109/PESTSE.2014.6805319
\bibitem{h93}
G. Hart, “Nonintrusive appliance load monitoring”, Proceedings of the IEEE. 80 (12): 1870-1891. 1992. doi:10.1109/5.192069
\bibitem{wz12}
Z. Wang, G. Zheng, “Residential Appliances Identification and Monitoring by a Nonintrusive Method”, IEEE Transactions On Smart Grid, Vol. 3, No. 1, pp. 80-92, March 2012, doi: 10.1109/TSG.2011.2163950
\bibitem{al17}
M. Aiad, P. H. Lee, “Non-intrusive monitoring of overlapping home appliances using smart meter measurements”, IEEE Conferences: Power and Energy Conference at Illinois (PEC), pp. 1-5, 2017, doi: 10.1109/PECI.2017.7935717
\bibitem{aasssr17}
S. R. Arrachman, M.F. Adiatmoko, A. Soeprijanto, M. Syai'in, M.S.A. Sidik, N.H. Rohiem, “Smart meter based on time series modify and extreme learning machine”, IEEE Conferences: 2nd International Conference on Automation, Cognitive Science, Optics, Micro Electro -Mechanical System, and Information Technology (ICACOMIT), pp. 86-92, 2017, doi: 10.1109/ICACOMIT.2017.8253392
\bibitem{h18}
J. R. Herrero, Á. L. Murciego, A. L. Barriuso, R. Carreira, “Non Intrusive Load Monitoring (NILM): A State of the Art”, 15th International Conference, PAAMS 2017 (pp. 125-138). Accessed: Mar. 29, 2019. [Online]. Available: \url{https://www.researchgate.net/publication/318510754_Non_Intrusive_Load_Monitoring_NILM_A_State_of_the_Art}
\bibitem{au17}
J. M. Alcalá, J. Ureña, Á. Hernández, D. Gualda, “Assessing Human Activity in Elderly People Using Non-Intrusive Load Monitoring”, Sensors, 2017, 17(2), 351. doi: 10.3390/s17020351
\bibitem{kk15}
J. Kelly, W. Knottenbelt, “Neural NILM: Deep Neural Networks Applied to Energy Disaggregation”. Sep. 28, 2015. Accessed: Mar. 29, 2019. [Online]. Available: \url{https://arxiv.org/pdf/1507.06594.pdf}
\bibitem{p15}
B. Packer, “7 reasons why utilities should be using machine learning”, March 24, 2015. Accessed: Mar. 29, 2019. [Online]. Available: \url{https://blogs.oracle.com/utilities/utilities-machine-learning}
\bibitem{bkb10}
D. Benyoucef, P. Klein, T. Bier, “Smart Meter with non-intrusive load monitoring for use in Smart Homes”, IEEE Conferences: International Energy Conference, pp. 96-101, 2010, doi: 10.1109/ENERGYCON.2010.5771810
\bibitem{vde10}
“Smart Energy 2020” in VDE Verband der Elektrotechnik Elektronik Informationstechnik e.V., Germany, 2010. Accessed: Mar. 29, 2019. [Online]. Available: \url{https://webuser.hs-furtwangen.de/~hoenig/2010/ETG.pdf}
\bibitem{m14}
S. Makonin, “Real-Time Embedded Low-Frequency Load Disaggregation”, PhD thesis, Simon Fraser University, School of Computing Science, 2014. Accessed: Mar. 29, 2019. [Online]. Available: \url{http://summit.sfu.ca/item/14410}
\bibitem{kw06}
R. Kahn, R. Wilensky, “A framework for distributed digital object services”, Corporation for National Research Initiatives, May 13, 1995. Accessed: Mar. 29, 2019. [Online]. Available:  \url{https://www.doi.org/topics/2006_05_02_Kahn_Framework.pdf}
\bibitem{sfi16}
“Digital identification of objects: technology and otherwise”.  Internet Support Foundation “Support Fund for Internet", Moscow, 2016, ISBN 978-5-9906425-4-6. Accessed: Mar. 29, 2019. [Online]. Available: \url{http://s.siteapi.org/808b25df7dd28c9.ru/docs/494d0aa0ed82f57b2dbd27a7b60c699d0148db30.pdf}
\bibitem{uspat18}
“A method and system for monitoring the parameters of the energy resources consumption process”, US Non-Provisional Application No.: 16/123,789. Sep. 6, 2018.
\end{thebibliography}
\end{document}